\documentstyle[
pre,
aps]{revtex}
\newcommand{\slaninafiginline}[1]{}

\def\slaninafigsize{ voffset=-400 hoffset=-60 
        hscale=80 vscale=80 }
\def\slaninafigspace{60mm} 
\begin{document}
\draft

\twocolumn[\hsize\textwidth\columnwidth\hsize\csname@twocolumnfalse\endcsname

\title{Capital flow in a two-component dynamical system}
\author{Franti\v{s}ek Slanina}
\address{Institute of Physics,
	Academy of Sciences of the Czech Republic,
	Na~Slovance~2, CZ-18221~Praha,
	Czech Republic\\
	and Center for Theoretical Study,
	Jilsk\'a~1,CZ-11000~Praha, Czech
	Republic\\
	e-mail: slanina@fzu.cz}
\author{Yi-Cheng Zhang}
\address{Institut de Physique Th\'eorique, Universit\'e de Fribourg,
P\'erolles, CH-1700 Fribourg, Switzerland} 
\maketitle

\begin{abstract}
A model of open economics composed of producers and speculators is
investigated by numerical simulations. The capital flows from the
environment to the producers and from them to the speculators. The price
fluctuations are suppressed by the speculators. When the
aggressivity of the speculators grows, there is a
transition from the regime with almost sure profit to a
very risky regime in which very small fraction of speculators have
stable gain. The minimum of price fluctuations occurs close to the
transition.
\end{abstract} 


\twocolumn]

\section{Introduction}
Economy is an intriguing complex dynamical system, understanding of which has
vital importance to the society
\cite{an_ar_pi_88,bou_ci_la_po_99,zhang_98,ausloos_98}. From the point of 
view of a physicist, it may be seen as a natural phenomenon, whose
microscopic ``laws of motion'' should be discovered and 
consequences drawn from them, amenable to experimental verification
(or falsification). Statistical physics is successfully involved in
investigation of various
collective dynamic phenomena, which come from interdisciplinary areas,
like car traffic \cite{na_pa_95},
city growth \cite{ma_zha_98a},
pedestrian dynamics \cite{he_sch_ke_mo_97},
forest fires \cite{dro_sch_92},
river networks \cite{ba_co_fla_gia_ma_ri_97}
or biological evolution \cite{pa_ma_ba_96}.

Like in these problems, numerical simulations of various ``minimal''
models of economic behavior
\cite{ta_mi_hi_ha_92,hi_ta_mi_ha_93,sa_ta_98,ba_pa_shu_97,sta_pe_98,bus_ri_99,ca_ma_zha_97,cha_zha_97,cha_zha_98,sa_ma_ri_99,jo_ja_jo_che_kwo_hui_98,ma_li_ri_sa_98,jo_ha_hu_98,jo_hu_jo_lo_98,ca_pla_gu_98,cavagna_99,ca_ga_gia_she_99,cha_ma_99,cha_ma_ze_99}
play important role, in parallel with 
analytical approaches  
\cite{cha_ma_ze_99,bou_po_97,co_bou_97,ta_sa_ta_97,eli_ko_98,bou_co_98,ga_bou_po_98,ma_zha_97a}. 
Even though the hope 
of detailed prediction of the future market behavior could be (at
least very probably) rarely satisfied, the knowledge of parameters,
which are crucial for the probabilistic properties of the economic
events is very important to the decision-makers on all levels.

One group of models investigated so far is based on threshold dynamics
of the players in the market
\cite{ta_mi_hi_ha_92,hi_ta_mi_ha_93,sa_ta_98} which was shown to be
equivalent to a stochastic process with both multiplicative and
additive noise \cite{ta_sa_ta_97,sa_ta_98} which then leads naturally
(see e. g.  \cite{drummond_92,so_co_97,sornette_97a,sornette_98c})
to the power law tails in the distribution of price changes, which are
observed in reality\cite{mantegna_91}.
Similar in spirit are the models based on a diffusion-annihilation
process, where buyers and sellers are considered as particles which
disappear once they meet \cite{ba_pa_shu_97,eli_ko_98}. The
percolation theory was invoked to account for the herd behavior of
market agents \cite{bou_po_97,co_bou_97,sta_pe_98,cho_sta_98}, which gives a
power law distribution truncated by an exponential, if the
connectivity of players is close but not exactly at the percolation
threshold. This is in accord with the ``truncated L\'evy''
distribution found in more refined recent analyses of stock market
data \cite{ma_sta_97a,bou_po_97,co_po_bou_97,bouchaud_98}.
A model based on non-linear Langevin equation was developed
\cite{bou_co_98} in order to explain the apparent ``phase transition''
character of market crashes, proposed recently by several authors.
(See e. g.  \cite{so_jo_bou_96,so_jo_97,va_au_bo_mi_98,so_jo_98,la_po_co_ag_bou_98,jo_so_99b}.)

Another model, which  
implements on-line adaptation of the community of players \cite{ca_ma_zha_97}
is able to reproduce very well also the scaling of price changes
\cite{co_po_bou_97,bouchaud_98,ma_sta_95,ga_ca_ma_zha_97}. 
More abstract approach is used in the minority game model
\cite{cha_zha_97,cha_zha_98,sa_ma_ri_99,jo_ja_jo_che_kwo_hui_98,ma_li_ri_sa_98,jo_ha_hu_98,jo_hu_jo_lo_98,ca_pla_gu_98,cavagna_99,ca_ga_gia_she_99,cha_ma_99,cha_ma_ze_99},
where real money is lacking. An important feature of this model is the
presence of a transition from chaotic to periodic phase, when the
number of players is increased. (It is quite interesting, that similar
transition seems to take place also in the threshold dynamics model
\cite{sa_ta_98}; when increasing the number of players, we may pass
from the intermittent behavior characterized by power-law tails to an
ordinary random-walk process. However, this phenomenon was not explicitly
investigated in \cite{sa_ta_98}.)

In our model, we want to take into account the fact, that
there are at least two types of investors. First, there are
individuals who product some commodity and need other commodities to
keep the  
production on. It is the latter type of economical subject, that  the market
was originally designed for. However, the second type of investor comes
soon, a speculator, who observes the price changes due to
disequilibrium between demand and offer, and makes profit from the
information carried by the price signal. However, it is expected, that
the influence of speculators is not at all purely negative. When they
discover some regularity in the price fluctuations, they make many of
it, but at the same time their activity has a feedback effect on the
price, so that the very fluctuations the speculators are exploiting are
destroyed. In fact, this may lead to overall decrease of price
fluctuations, making the market more stable.

The question is, whether this common sense reasoning can be supported
by more rigorous arguments. The scope of the present work is to
implement a model of market, in which the mutual influence of
producers and speculators may be studied.

\section{Description of the two-component model}
Economy is an open system. Like thermodynamic systems can
self-organize into a low-entropy state only under condition that there
is flow of energy through the system, non-trivial
self-organization in the market is driven by the supply of wealth
from the surrounding environment. So, first fundamental players in the
economic 
game are producers, which exploit the outside opportunities. In our
model, their real economic activity will be mimicked by buying and
selling a single commodity in a regular manner. For simplicity, we
will call it 
stock. Each producer is supplied periodically fixed amount of stock
and money from the outside. 

Real market needs also speculators, which play a positive role in
absorbing temporary disequilibrium of demand and offer. The result is
a more liquid market with reduced fluctuations. The speculators are
selfish, but without any explicit wish, besides the net gain. We may
look at them as providers of a service, for which they are
paid. However, none of them provides individually a specific service,
but their utility stems from a collective effect.

An important feature of our model is the possibility of both producers
and speculators to abstain from the game, if they feel that it does
not pay to participate. So, the number of players in each group is
self-adjusted, depending on the parameters of the model. This is
similar in spirit to the grand-canonical ensemble in statistical physics.

Borrowing the biological terminology, the producers are ``autotrophs''
who live in a symbiosis with ``heterotrophic'' speculators. Like
in biological communities, we implement here also Darwinian evolution
of the speculators, so that they adapt collectively to the actions of
producers. 

Let us be more specific now.
The dynamical system we are going to investigate is a simplified model
of a market, in discrete time. In each step, some
amount of stock is traded. The price of stock $x(t)$
as a function of time $t$ is the output signal of the market. The
price is the manifestation of large amount of ``microscopic'' activity
due to players on the market. Each player is characterized by two
dynamical variables, the amount of stock $S_i(t)$ and the
amount of money $B_i(t)$, where index $i$ denotes the player.
So, the total capital owned at time $t$ by $i$'th player is
$W_i(t)=B_i(t)+x(t)S_i(t)$.

There are two kinds of players in our model. There are $N_{\rm p}$
producers and $N_{\rm s}$ speculators. 
We denote  $N = N_{\rm p} + N_{\rm s}$ the total number of players. The
producers follow a fixed strategy of buying and selling,
irrespective of the current or past price. On the other hand, the
decisions of speculators are based on the analysis of the past
evolution of price. 

The strategies are characterized as follows.
Each producer, $i=1,2,...,N_p$ has its own period $\tau_i$ and
time-scale $T_i$ on which he or she invests. The periods are chosen
randomly among numbers 2 to 6, the time-scales randomly from 7 to 10.
The investment follows a
random but quenched pattern $a_i(t^\prime)$,
$t^\prime=0,1,...,\tau_i-1$. In order 
to avoid a systematic excess on the demand or offer side, we require
that the investment is balanced for each producer, $\sum_{t^\prime =
0}^{\tau_i-1}a_i(t^\prime) = 0$. Apart from this constraint, the $a_i$'s
are drawn from uniform distribution on the interval $(-1,1)$ 
Finally, all producers are given the same overall amplitude of their
investment $\epsilon$.  
At the time step $t$, the producer $i$ 
participates, if he or she has positive capital. In this case, he or
she attempts to
buy the following amount of the commodity
\begin{equation}
\overline{\Delta S_{i}} = 
\epsilon (a_i([t/T_i]  \;{\rm mod}\; \tau_i) 
	- \lambda \ln {x(t)\over <W>})
\; .
\label{eq:dsprod}
\end{equation}
In this formula, we denote by $[t/T_i]$ integer division of $t$ by
$T_i$ and 
$<W>=\frac{1}{N}\sum_{i=1}^N W_i$ is the average wealth of the
players. The last term with the logarithm 
expresses the fact, that the stock has its intrinsic value. Its
price is measured relatively to the average wealth of the population,
so that if the price is larger, the strategy of the producer is
slightly biased towards selling, while if the price is lower, the
producer is more likely to buy. The use of logarithm follows from the fact,
that the evolution of price is a multiplicative process, rather than
additive one. Analogical term plays a crucial role
also in the analytic approach of Ref. \cite{bou_co_98}. 
The parameter $\lambda$ measures the
strength of the bias caused by the intrinsic value of the
commodity. Throughout this article, we use the value $\lambda=0.01$.

The speculators differ from the producers in two aspects. First, they
do not feel the intrinsic value of the traded product, so that the
logarithmic term in the Eq. (\ref{eq:dsprod}) is missing. But the
crucial difference resides in the ability of speculators to analyze
the past price signal and decide according to their expectations about
the future.
As the producers do, the speculators may have their time-scales on which
they analyze the signal and also different memory length. However, in
the present work we limit 
ourselves only to the case when all speculators have their time-scale
equal to one step of the dynamics and memory is fixed and uniform in
the whole community of speculators.

 The speculators have memory $M$. It means that they are able to use
information of $M$ previous values of price,
 $x(t),x(t - 1),...,x(t - M)$. This sequence
is then transformed in a bit string
$\sigma=[\sigma_1,\sigma_2,...,\sigma_M]$ containing the information
whether the price went up or down in a given instant in the past. We
adopted the convention that 0 means increase and 1 means decrease
of the price. Therefore $\sigma_j=\theta(x(t-j)-x(t-j+1))$, where
$\theta(x)$ is the Heaviside function.
The strategy of the $i$-th speculator is the function which prescribes
for each bit string, whether the speculator should buy or sell. We
adopt the convention, that 0 means selling and 1 buying. Then the
strategy is a function $\Sigma_i(\sigma)$ with possible values 0 or 1.

The strategy has a score $b_i(t)$ counting its
success rate at time $t$. 
If it predicted correctly the change of the price from step $t$ to
$t+1$ one point is added to the score, otherwise one point is subtracted.
There is in principle a non-trivial question, what we mean by saying
``predicted correctly the price change''. In fact, it should be
checked {\it a posteriori } that the rule we used for distinguishing
successful strategies corresponds really to winning behavior. However,
in our model, we used a prescription for price change, which enables us
to say {\it a priori } which strategy did a good job.
At this moment, we present the rule of success rate counting and
return to the justification of this rule later, when we will speak of the
prescription for price change. The essence is, that it is good to buy
if the price will go down and vice versa.
So, when the player is inserted in the market, its score is set to
zero, and in each successive step, the score is updated as follows:
$b_i(t+1) = b_{i}(t) + 1$ if $\theta(x(t)-x(t+1)) = \Sigma_{i}(\sigma(t))$
and $b_{i}(t+1) = b_i(t) - 1$ in the opposite case.

There is a Darwinist selection among speculators. Each 5 steps, the
speculator with lowest capital is removed and replaced by new player
with newly chosen strategy. However, the capital, amount of stock and
money is inherited from the removed player. This amounts to not really
remove the player, but rather the player picks new strategy instead of
the old doomed one. We define the age $v_i(t)$ of speculator $i$ as
number of time steps since the last replacement of the strategy.
We implemented also random mutations, which affect equally good and
bad players. Each 57 steps a player is chosen at random and its 
strategy is randomly changed. 

If the speculator feels, that the strategy is bad, he or she may
abstain, in order to avoid losses. For the player $i$ to participate,
we require that $b_i(t)/v_i(t) > 0.05$.

Those speculators, who do participate, attempt to buy the following
amount of stock:
\begin{equation}
\overline{\Delta S_{i}} = \delta (2\Sigma_i (\sigma(t)) -1)\;\; .
\label{eq:dsspec}
\end{equation}

When we know what amount of stock the players want to buy or sell, we
can compute the change of price. It is not clear {\it a priori } what
precisely the price change should be. The only obvious requirement is,
that the price should go up, when there is more demand than offer, and
go down, when the offer prevails. 
The demand is $D=\sum_{i,\overline{\Delta S_{i}}>0}\overline{\Delta S_{i}}$
and the offer $O=-\sum_{i,\overline{\Delta S_{i}}<0}\overline{\Delta
S_{i}}$. In the previous works, two recipes for the price change were
used. In \cite{ca_ma_zha_97} the time averages of demand and offer
were computed and new price was obtained by multiplying the old price
by the ratio of average demand to average offer. Essentially the same
prescription is applied in \cite{bou_co_98}.

On the other hand, within the approaches based on threshold dynamics or
diffusion-annihilation processes \cite{sa_ta_98,ba_pa_shu_97} the
deal is realized when the bid and ask prices meet, which determines
the reported stock price at that moment. (It is interesting to note,
that between the deals the price is undefined.)

Here we adopt an approach closer to the former one. The new price is
computed by multiplying the old price by a factor, which increases
with the ratio $D/O$.
\begin{equation}
x(t+1) = F({D\over O})x(t)
\label{eq:xt1}
\end{equation}
where the function $F(r)$ should obey two conditions, $F(1)=1$ and
$F^\prime(r)>0$. The simplest choice consists in taking $F(r)=r$, but
as we have seen in our simulations, this leads to price fluctuations
far beyond realistic values. So, we use a non-linear form which
suppresses the fluctuations,
\begin{equation}
F(r) = \exp(\alpha \tanh(\log(r)/\alpha))
\label{eq:F}
\end{equation}
which has the property that $F(r)\simeq r$ for $r$ close to 1.
We have found that $\alpha = 0.02$ gives realistic price fluctuations,
so we keep this value throughout the simulations.

There should be conservation of stock in each trade event. Therefore
the amount actually traded by a single player is not the same as the
attempted volume.
If $D>O$ then if $\overline{\Delta S_{i}}>0$, actual change of stock
is lower than attempted, $\Delta S_{i} = \frac{O}{D}\overline{\Delta
S_{i}}$. If $\overline{\Delta S_{i}} <0$ the actual traded amount is
the same as attempted, $\Delta S_{i} = \overline{\Delta S_{i}}$.

If on the contrary $D<O$ then if $\overline{\Delta S_{i}}<0$, $\Delta
S_{i} = \frac{D}{O}\overline{\Delta 
S_{i}}$, if $\overline{\Delta S_{i}} >0$, $\Delta S_{i} =
\overline{\Delta S_{i}}$. 

After the trade is completed, the new amount of capital, stock and
money is
\begin{eqnarray}
W_i(t+1) &=& B(t) + x(t+1)S_i(t)
\label{eq:wt1}\\
S_i(t+1) &=& S_i(t) + \Delta S_i
\label{eq:st1}\\
B_i(t+1) &=& B_i(t) - x(t+1) \Delta S_i
\label{eq:bt1}
\end{eqnarray}

Finally, each 120 steps, the producers receive wealth from
outside. The value of the influx is governed by the parameter
$\eta$. The total amount distributed among producers is $\eta N_{\rm
p}$, but who do not participate in that moment, does not receive
anything. If $p_{\rm p}$ is the fraction of currently participating
producers, those who do participate, increase their capital by $\eta /
p_{\rm s}$.

Summarizing the algorithm which defines our model, the following
operations are performed in each step.
\vspace*{3mm}

{\it (i)} Calculate amount of the commodity, attempted to buy by
producers (Eq. (\ref{eq:dsprod})) and speculators (Eq. (\ref{eq:dsspec})).

{\it (ii)} Calculate demand and offer and from them the new price,
according to (\ref{eq:xt1}) and (\ref{eq:F}).

{\it (iii)} Calculate new values of capital, stock and money according
to (\ref{eq:wt1}), (\ref{eq:st1}), and (\ref{eq:bt1}).
\vspace*{3mm}

The following actions are performed periodically.
\vspace*{3mm}

{\it (iv)} If the step is a multiple of 57, randomly chosen speculator
changes randomly its strategies. If the step is a multiple of 120,
wealth is added to producers.
\vspace*{3mm}

An important property of the above algorithm is that the game has {\it
minority} character. Indeed, because the price is established after the attempted
traded amounts $\overline{\Delta S_{i}} $ were fixed, it is an
advantage to sell, if the price goes up and buy, when the price comes
down. Therefore this means that going counter the
majority is an. This enables us to decide, what score should be attributed
to the strategies: those leading to minority side receive +1 point,
those which lead to majority side, receive -1 point. In the abstract
minority game \cite{cha_zha_97,cha_zha_98} it was possible only for
less than half players to be on the minority side, of course. The
presence of producers, however, makes it possible for any number of
speculators to have minority strategy, at least in principle.

\section{Results}

\begin{figure}[hb]
  \centering
  \vspace*{\slaninafigspace}
  \includegraphics{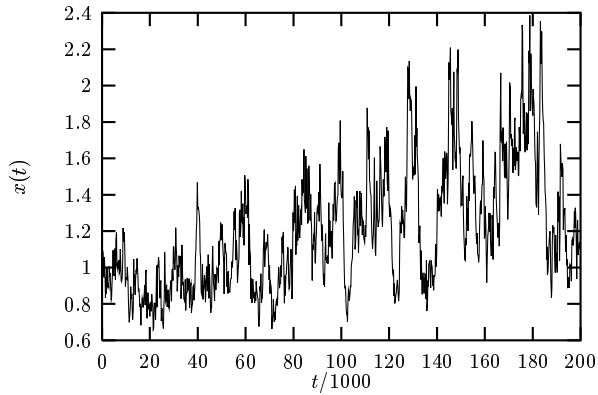}
  \caption{Price averaged over 10 steps, for for $N_{\rm p} = 50$, 
  $N_{\rm s} = 1000$, $M=10$, $\epsilon = 0.01$, $\delta=0.001$,
  $\eta=0.001$.}
  \label{fig:price10}
\end{figure}

The fundamental variable of our model is the price. A typical time
evolution of price is shown in Fig. \ref{fig:price10}. The long-time
average of price grows, due to the fact, that capital is regularly
injected into the system. We observed, that the increase of price is
higher if the capital influx is higher. We measured also the price
fluctuations. We observed, that the relative price fluctuations remain
constant in the long-time average, so that the absolute fluctuations
grow with time with the same rate as the long-time average of price does.

One of the main questions is, how the relative price fluctuations is
changed by the presence of the speculators. 
We define the the time-averaged relative price fluctuations using
exponential averages \cite{ca_ma_zha_97}
\begin{equation}
f=\sqrt{{\sum_{t=1}^{T}\lambda^{T-t+1}x^2(t)
\sum_{t=1}^{T}\lambda^t
\over\left(\sum_{t=1}^{T}\lambda^{T-t+1}x(t)\right)^2}-1}
\end{equation}
where $T$ is duration of the simulation run and the parameter was
chosen $\lambda=0.9999$.

\begin{figure}[hb]
  \centering
  \vspace*{\slaninafigspace}
  \includegraphics{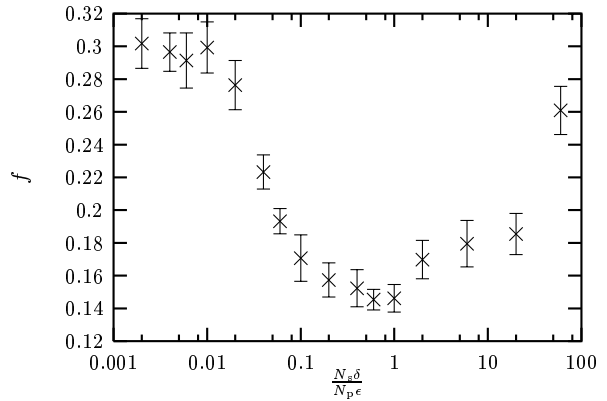}
  \caption{Relative price noise for $N_{\rm p} = 50$, 
  $N_{\rm s} = 1000$, $M=10$, $\epsilon = 0.01$, $\eta=0.001$
  Duration of the run is $10^5$. Data are averaged over 10 independent
  runs.}  
  \label{fig:avall-noise}
\end{figure}

The relative weight of the
speculators compared with the producers is the quantity $\xi = N_{\rm
s}\delta/N_{\rm p}\epsilon$. Figure \ref{fig:avall-noise} shows the
dependence of 
the time-averaged relative price fluctuations
 on $\xi$. We can see a pronounced minimum
around $\xi = 0.5$, which suggests, that from the point of view of the
price stability, there is an optimal weight of the speculators. 

\begin{figure}[hb]
  \centering
  \vspace*{\slaninafigspace}
  \includegraphics{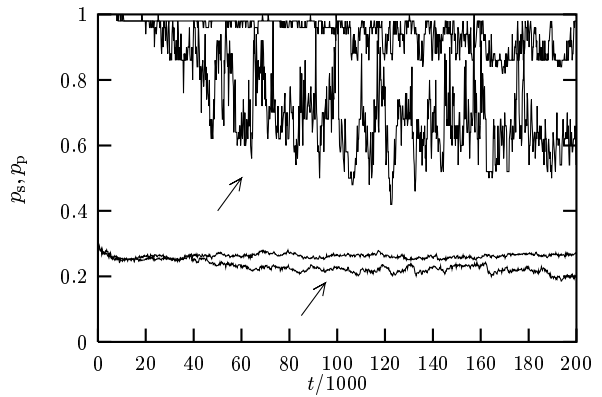}
  \caption{Time dependence of the participation of producers (upper
  pair of curves) and speculators (lower pair of curves), for
  $N_{\rm p} = 50$,    $N_{\rm s} = 1000$, $M=10$, $\epsilon =
  0.01$, 
  $\delta=0.0001$ and two values of $\eta=0.005$ and
  $\eta=0.001$. The arrows 
  are pointing to the curves corresponding to $\eta=0.001$.}
  \label{fig:partic-time}
\end{figure}

\begin{figure}[hb]
  \centering
  \vspace*{\slaninafigspace}
  \includegraphics{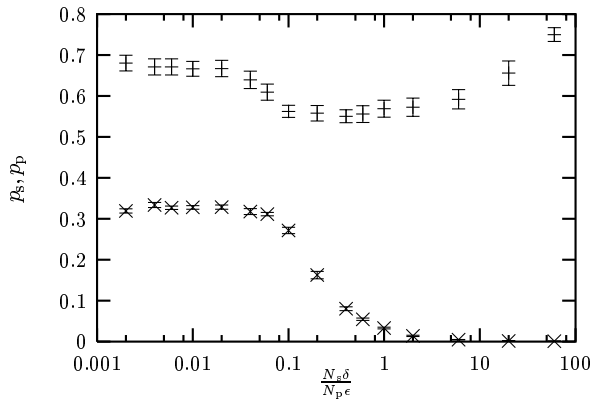}
  \caption{Average participation of producers ($+$) and
  speculators ($\times$) for $N_{\rm p} = 50$, 
  $N_{\rm s} = 1000$, $M=10$, $\epsilon = 0.01$,
  $\eta=0.001$. Duration of the run is $10^5$ steps.
  Data are averaged over 10 independent runs.}  
  \label{fig:avall-partic}
\end{figure}

This phenomenon can be better understood, when we observe the
participation of producers ($p_{\rm p}$) and speculators ($p_{\rm
s}$), defined as the percentage of those, 
who take part in the trading. The Fig. \ref{fig:partic-time} shows
the time dependence of the participation in a typical run. 
After a
transient period, the participation fluctuates around a stationary
value, which grows with the capital influx $\eta$. The dependence of
the time-averaged participation on the parameter $\xi$ is shown in the 
Fig. \ref{fig:avall-partic}. The most important observation is the
substantial decrease of the participation of the speculators for the
value of $\xi$ 
close to $0.5$. The participation of producers has a shallow minimum
around the same value $\xi=0.5$, which is also close to the position
of the minimum in the relative price noise.

The following picture arises from these observations. If the
speculators are too prudent (small $\delta$), the price fluctuations
are high, because of the demand-offer disequilibrium. The price
changes follow a periodic pattern induced by the periodic quenched
pattern of trading of each individual producer. Speculators are able
to extract the information about the periodic price changes and use it
to make profit. Because they trade with little capital (low
$\delta$) they do not influence much the price and many speculators
can gain. On the other hand, the gain is also small. 

If the investors became more aggressive, by increasing $\delta$, they
have larger influence on the price, which leads to the suppression of
the price fluctuations, but at the same time their ability to use the
periodic price fluctuations to make profit is also suppressed. This
results in the fact, that less speculators have successful strategy
and less speculators participate. 

There is a transition between the
low-aggressivity regime, where many speculators make little profit, and
the high-aggressivity regime, where few speculators can make large
profit. The transition occurs around $\xi=0.5$ and it is characterized
by optimum of the relative price noise and also by the minimum of the
participation of the producers, which means that less producers are
able to gain. The advantage is more stable
price, the disadvantage is less profit for the producers. This is a
manifestation of the common sense consideration, repeated in all
economics literature, stating that higher profit is more risky.

\begin{figure}[hb]
  \centering
  \vspace*{\slaninafigspace}
  \includegraphics{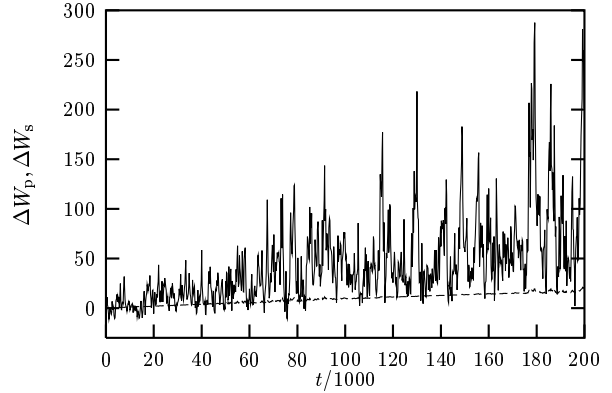}
  \caption{Total increase of wealth of producers (full line) and
  speculators (dashed 
  line) time-averaged over 120 steps, for
  for $N_{\rm p} = 50$,  
  $N_{\rm s} = 1000$, $M=10$, $\epsilon = 0.01$, $\delta=0.00001$,
  $\eta=0.001$.}
  \label{fig:wa-00001}
\end{figure}
\begin{figure}[hb]
  \centering
  \vspace*{\slaninafigspace}
  \includegraphics{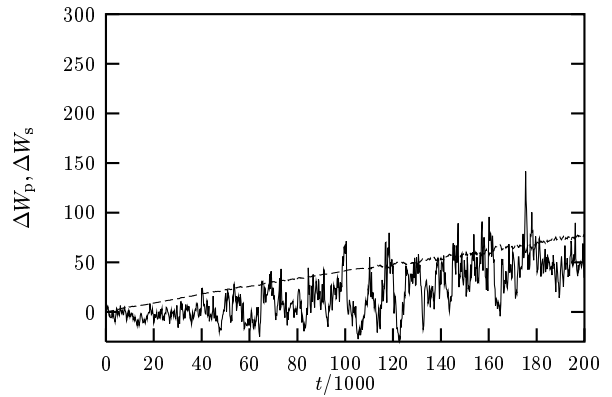}
  \caption{Total increase of wealth of producers (full line) and
  speculators (dashed 
  line) time-averaged over 120 steps, for
  for $N_{\rm p} = 50$,  
  $N_{\rm s} = 1000$, $M=10$, $\epsilon = 0.01$, $\delta=0.0001$,
  $\eta=0.001$.}
  \label{fig:wa-0001}
\end{figure}
In
Figs. \ref{fig:wa-00001} and \ref{fig:wa-0001} the situation is
illustrated by time 
evolution of the increase of the 
total wealth of producers and speculators, defined by
\begin{eqnarray}
\Delta W_{\rm p}(t)=&\sum_{i\in{\rm producers}}(W_i(t)-1)\\
\Delta W_{\rm s}(t)=&\sum_{i\in{\rm speculators}}(W_i(t)-1)
\;\; .
\end{eqnarray}
Lower $\delta$
(Fig. \ref{fig:wa-00001}) is characterized by large fluctuations of
the capital of the producers, while capital of the speculators grows
slowly. When the $\delta$ is larger (Fig. \ref{fig:wa-0001}), the
capital of producers fluctuates less, but the speculators have
significantly larger profit.

\begin{figure}[hb]
  \centering
  \vspace*{\slaninafigspace}
  \includegraphics{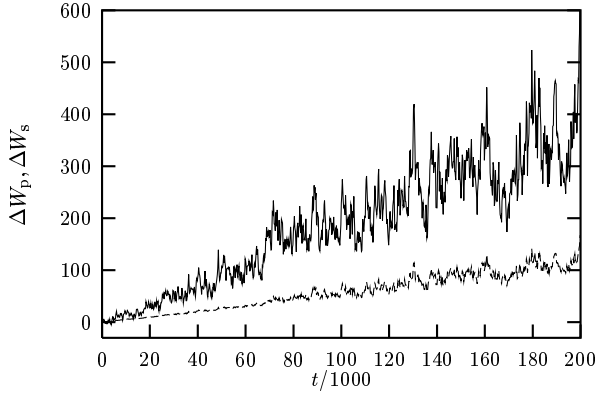}
  \caption{Total increase of wealth of producers (full line) and
  speculators (dashed 
  line) time-averaged over 120 steps, for
  for $N_{\rm p} = 50$,  
  $N_{\rm s} = 1000$, $M=10$, $\epsilon = 0.01$, $\delta=0.0001$,
  $\eta=0.005$.}
  \label{fig:wa-0001-005}
\end{figure}
The influence of the influx of capital into the system, measured by
the parameter $\eta$, can be seen by
comparing the Figs. \ref{fig:wa-0001} and \ref{fig:wa-0001-005}.
The picture remains the same qualitatively, however larger influx
means larger profit preferably for the producers, while the increase
of the profit of the speculators is much lower. The
Fig. \ref{fig:partic-time} shows, that the participation of the
producers is much more influenced by the parameter $\eta$ than the
participation of the speculators.

\section{conclusions}

We introduced and studied a model of open economics. The influx of
capital leads to the coexistence of producers and speculators. 
We showed, that the presence of the speculators can be useful to the
economics, by suppression of the price fluctuations. If we increase the
aggressivity of the speculators, there is a smooth transition from the
regime with small, but less risky profit for the speculators, to the
regime with larger profit, but accessible to smaller fraction of the
set of the speculators. The transition occurs close to the minimum of
the price fluctuations, which is the optimal state for the
producers. If we accept the supposition, that the optimal strategy for
the speculators should be derived by a compromise of the mutually
exclusive requirements of risk and profit, we can conclude, that the
optimum for the speculators lies also in the transition region. As a
result, the optima for producers and speculators lie close one to the
other, and their mutual coexistence should be better described as
symbiosis than parasitism.

\acknowledgments{
We acknowledge the support from the European TMR Network-Fractals
c.n. FMRXCT980183.
F.S. wishes to thank the University of Fribourg,
Switzerland, for the 
financial support and kind hospitality.}


\end{document}